# Thermal skyrmion diffusion applied in probabilistic computing

Jakub Zázvorka[a], Florian Jakobs[b], Daniel Heinze[a], Niklas Keil[a], Sascha Kromin[a], Samridh Jaiswal[a,c], Kai Litzius[a,d,e], Gerhard Jakob[a], Peter Virnau[a], Daniele Pinna[a], Karin Everschor-Sitte[a], Andreas Donges[b], Ulrich Nowak[b], Mathias Kläui[a,d]

[a] Institut für Physik, Johannes Gutenberg-Universität Mainz, DE-55099 Mainz, Germany

[b] Fachbereich Physik, Universität Konstanz, Universitätsstraße 10, DE-78457 Konstanz, Germany

[c] Singulus Technologies AG, DE-63796 Kahl, Germany

[d] Graduate School of Excellence Materials Science in Mainz, DE-55128 Mainz, Germany

[e] Max Planck Institute for Intelligent Systems, 70569 Stuttgart, Germany

## Abstract

Thermally activated processes are key to understanding the dynamics of physical systems. Thermal diffusion of (quasi-)particles for instance not only yields information on transport and dissipation processes but is also an exponentially sensitive tool to reveal emergent system properties and enable novel applications such as probabilistic computing. Here we probe the thermal dynamics of topologically stabilized magnetic skyrmion quasi-particles. We demonstrate in a specially tailored low pinning multilayer material system pure skyrmion diffusion that dominates the dynamics. Finally, we analyse the applicability to probabilistic computing by constructing a device, which uses the thermally excited skyrmion dynamics to reshuffle a signal. Such a skyrmion reshuffler is the key missing component for probabilistic computing and by evaluating its performance, we demonstrate the functionality of our device with high fidelity thus enabling probabilistic computing.

## Introduction

Magnetic skyrmions are localized and topologically stabilized magnetic textures with great potential for applications in data storage[1], logic[2], including synaptic devices[3] and probabilistic computing,[4] and further spintronic devices[5,6]. Skyrmions can be stabilized by the chiral Dzyaloshinskii-Moriya exchange interaction (DMI)[7,8] in systems with bulk inversion asymmetry such as B20 compounds[9–12] and thin film multilayers with structural inversion asymmetry[13,14]. In magnetic thin films skyrmions can occur naturally as thermodynamically stable phases, but in particular by engineering multilayer stacks they can exist also as metastable structures, which can be selectively written, displaced and deleted[14]. The interaction potentials have so far been studied primarily theoretically and are found to be repulsive at short distances[2,15] leading to possibly good stability. For the displacement, efficient motion by spin transfer torques at ultra-low current densities[16,17] and high speeds due to spin-orbit torques[18,19] has been reported including highly reproducible dynamics that can be measured with pump-probe techniques[20].

While this favourable combination of properties has raised expectations for using skyrmions in devices, most studies, in particular of the dynamics, theoretically consider only the 0 K limit and also experimentally thermal effects on the dynamics have hardly been studied. However, thermally induced skyrmion dynamics can be used for unconventional computing approaches as they have been predicted to be useful for probabilistic computing[4]. Diffusive skyrmion motion is considered in a signal reshuffling device as part of skyrmion-based probabilistic computing architecture, thus making it highly desirable to realize and study thermal effects in experimentally accessible skyrmion systems.

In a first theoretical study, recently Schütte et al.[21] exploited a generalized Thiele equation[22] to investigate the diffusion dynamics of skyrmions. They derived an expression for the diffusion coefficient, which turned out to be proportional to the Gilbert damping coefficient $\alpha$. This surprising deviation from the normal theory of Brownian particle motion, where the diffusion coefficient is inversely proportional to the friction, led the authors to the conclusion that thermal diffusion is strongly suppressed for skyrmions in magnetic systems where the damping is typically small. Based on this conclusion, current common wisdom dictates that the skyrmion position should be stable and skyrmion motion should solely follow any time-dependent current. The quantitative appropriateness of the conclusions is however a priori not clear and has to date not been tested, neither experimentally nor by atomistic spin model simulations.

In our work, we uncover thermal diffusive skyrmion dynamics by a combined experimental and numerical study and employ the observed thermal motion to realize a skyrmion based device used for reshuffling input data streams as the key missing component for probabilistic computing. We develop special ultra-low pinning multilayer stack systems, where we can realize diffusive skyrmion motion at room temperature that dominates the dynamics on the experimentally accessible timescales. By a comparison with finite temperature computer simulations based on the stochastic Landau-Lifshitz-Gilbert (LLG) equation of motion for an atomistic spin model, we show that our findings can be explained by diffusion in a non-flat energy landscape. We apply thermal skyrmion diffusion in a skyrmion based signal reshuffler device, which is the necessary component for probabilistic computing. We assess the device performance and find good decorrelation and signal fidelity thus facilitating probabilistic computing using skyrmions.

## Results

We investigate skyrmions in stacks with structural inversion asymmetry based on Ta/Co$_{20}$Fe$_{60}$B$_{20}$/Ta/MgO/Ta (for details, see Methods section). The Magneto-optical Kerr-effect (MOKE) hysteresis loops of the used sample shows a typical hour-glass shape, usually related to a skyrmionics material [19,23]. By varying the OOP field, we can choose the skyrmion density and radius. By current injection (for details, see Methods section), we can controllably nucleate single skyrmions, thereby controlling the density of skyrmions precisely so as to avoid skyrmion-skyrmion interactions.

We first study the skyrmions present for a fixed perpendicular field value of 0.35 mT leading to a skyrmion radius of approx. 0.9 μm. To understand their topology, we analyse spin-orbit torque driven displacements. By analysing the displacement with and without in-plane fields, we conclude that the skyrmions studied here have a topological charge of Q= 1 [18]. After having injected the desired number of skyrmions into a device, we then observe skyrmions in real-time with the Kerr microscope without any further current injection or any other external stimulus. In contrast to previous reports and our previous results in different stacks[19,20], we here clearly observe skyrmion motion after the system has relaxed and no current pulse after-effect can be present anymore (for a movie, see Suppl. Video 1). At constant conditions (no current, constant temperature, constant field) the skyrmions move randomly throughout the sample. To evaluate this skyrmion motion, individual skyrmions were tracked and an example of several typical skyrmion trajectories is shown in Fig. 1.

Having established motion without any external artificial excitations, the first step is to identify the origin of the dynamics. One possible origin is the presence of thermal effects that can lead to diffusion from the sub- to the super-diffusive regime and, in the simplest case, to pure diffusion. The different regimes can be identified from the mean squared displacement (MSD) that we measure for different skyrmions. For pure diffusion of Brownian motion assuming rigid particles with no correlations, the average displacement should be proportional to the square root of the time. We thus plot the MSD

$$\langle \left(R_v(t) - R_v(0)\right)^2 \rangle = 2\mathfrak{D} t \cdot d \tag{1}$$

that should then be proportional to the time $t$ elapsed and the Diffusion coefficient $\mathfrak{D}$. The parameter $d$ denotes the dimension of the system and for our measurement has the value of 2. The resulting MSD as a function of time at room temperature averaged for a number of skyrmions is shown in the inset in Fig. 1. We find an extremely good agreement with a linear dependence using the Eq. (1) and a linear fit of the evaluated data reveals the diffusion coefficient of the material stack to be $\mathfrak{D} = 0.31(15) \times 10^{-12} \text{m}^2\text{s}^{-1}$ at a temperature $T = 296$ K.

With the MSD being linearly proportional to time, we therefore have as a first key result that skyrmions exhibit diffusive dynamics. This is in line with the analytical predictions[21] and we also confirm this linear dependence in numerical simulations, see Fig. 2 (for details on the simulations, see Methods section).

We now compare the diffusion properties to the theoretical models. The theoretical prediction for the diffusion coefficient

$$\mathfrak{D} = k_B T \frac{\alpha \mathcal{D}}{(\alpha \mathcal{D})^2 + \mathcal{G}^2} \tag{2}$$

for a rotationally symmetric skyrmion has previously been derived by Schütte et al.[21]. The gyrocoupling $\mathcal{G} = 4\pi Q M_S/\gamma$ and the trace over the dissipative tensor $\mathcal{D} = M_S/(2\gamma) \int (\partial_x \mathbf{S})^2 + (\partial_y \mathbf{S})^2 d^2 r$ [22] are hereby independent of the skyrmion's size, hence a comparison between the skyrmions in the Ta/Co$_{20}$Fe$_{60}$B$_{20}$/Ta/MgO/Ta stack used for our measurements and a much simpler theoretical model is sufficient to gain the necessary understanding of the observed diffusion dynamics. To numerically obtain results of the simulations with realistic computational effort, our simulations use a model for skyrmions with properties similar to those found for instance in a Pt$_{0.95}$Ir$_{0.05}$/Fe/Pd(111) ultrathin film containing a single Fe monolayer[15,24]. This system is chosen as it has been extensively studied in past theoretical works by Rózsa et al.[15,24] using a robust multiscale model of ab initio calculations and Langevin dynamics simulations (for details, see Methods section).

In line with the experimental data, we find in Fig. 2 for the MSD of the simulated skyrmion trajectories a diffusive skyrmion motion where the MSD increases linearly with time. For a classical particle that satisfies the Einstein-Stokes relation, the MSD is expected to increase when the friction is decreased. However, skyrmions show the opposite behaviour and the diffusive motion increases with the friction $\alpha \mathcal{D}$. In most materials $\alpha$ is much smaller than unity while $\mathcal{D} \sim \mathcal{G}$ have the same magnitude. Therefore, the reason for this unexpected behaviour lies in the gyrocoupling $\mathcal{G}$ which suppresses the diffusive motion of skyrmions independently of $\alpha$. Analyzing the MSD with Eq. (1), we calculate the diffusion coefficient from simulations at $T = 2$ K for different values of the damping constant $\alpha$, displayed in Fig. 3. While our simulations confirm the validity of Eq. 2, the conclusions drawn by Schütte *et al.* regarding the irrelevance of diffusive skyrmion motion are not confirmed. Quantitatively, our findings rather stress the importance of diffusive skyrmion dynamics on experimentally relevant length and times scales.

Beyond analysing the intrinsic system properties, thermal skyrmion dynamics has also been put forward as a mechanism that allows for realizing the key missing component for probabilistic computing using skyrmions. Instead of performing mathematical operations via precise bitwise operations, probabilistic (stochastic) computing seeks to operate on random sequences of bits which encode definite quantities in the overall ratio of "1"s and "0"s. To realize for instance a calculation of a product of two numbers, probabilistic computing needs an AND gate, which has been established for skyrmions in Ref. [25]. However, logic gates used in probabilistic computing circuits require uncorrelated inputs for proper operation. Correlated inputs result in unwanted propagations of correlations which ultimately lead to incorrect computational results (for details, see Figure 2 in Ref. [4]).

For a proper operation, input signals have to be periodically reshuffled while preserving the p-value (signal up/down ratio) used for the actual computing step[26]. Harnessing our observed naturally diffusive skyrmion motion and particle number conservation due to the skyrmion stability poses promising strategy for the development of a device. A first proof-of-concept can be performed by sampling telegraph noise signals and de-correlating them[4]. To implement the reshuffler concept, we thus need to drive skyrmions through a reshuffling chamber and generate an output signal that is uncorrelated to the input while maintaining the same p-value with high fidelity.

We pattern the investigated material into the reshuffler design as shown in Fig. 4. To study the skyrmion reshuffler operation, we compare independent measurements for the respective "0-bit" and "1-bit" channels (see Fig. 4). Skyrmions are nucleated at the contact and then driven through the reshuffling chambers by applying a DC current. Due to the stability of the skyrmions in our system, the number of skyrmions stays constant (keeping the p-value of the input also for the output) while not counting in pinning and annihilation effects. The thermal diffusion leads to a reshuffling of the bits (decorrelation of the output and input streams). To assess the functionality of the device, we thus compare the p-value of the input and output to determine the fidelity. The deterministic generation of skyrmions is engineered to be an unrelated process. By continuously generating skyrmions at the contact by DC current and driving them through the reshuffler chamber we measure input and output streams by detecting each skyrmion that crosses the blue and orange threshold lines (Fig. 4). Upon crossing the blue threshold line in front of the reshuffling chamber, the corresponding bit is triggered and the input signal gains its value for the time until the other bit is triggered when a skyrmion passes the threshold line in the other chamber. The output signal is obtained using the same principle, by skyrmions crossing the orange threshold line depicted in Fig. 4. We then record and evaluate the full operation as shown in Fig. 4. A real-time movie of the skyrmion reshuffler operation and the signal evaluation is shown in Suppl. Video 2. The p-value of the incoming and outcoming signal is calculated as their mean value, while the correlation of the two gained signals is evaluated using the equation for Pearson correlation factor $\rho$

$$\rho = \frac{cov(In,Out)}{\sigma_{In} \cdot \sigma_{Out}}, \tag{3}$$

where $cov$ means the covariance of the signals and $\sigma$ is the standard deviation of the respective signals. Evaluating the skyrmion reshuffler operation using a current density of $j = 3 \times 10^8 \, Am^{-2}$ we obtain the p-value to the input as $p_{input} = 0.51 \pm 0.08$. We find a negligible change of the p-value between input and output signals, $\Delta p = 0.01 \pm 0.08$, showing the high fidelity of the signal retention. The calculated correlation factor has the value $\rho = 0.11 \pm 0.14$ which denotes a generation of an output signal that is highly uncorrelated to the device input signal.

In conclusion, we identify pure diffusion of the skyrmions from the observed linear dependence of the mean square displacement with time. This is in qualitative agreement with analytical theory[21] and numerical simulations. However, we find that in contrast to the conclusions drawn earlier[21] the skyrmion position is thermally unstable and its diffusive motion actually dominates the dynamics in our low pinning system.

Having established the strong thermal dynamics, we show that this opens up a broad range of powerful statistical mechanics analysis methods to skyrmions. As an example, we use the thermal dynamics to spatially-resolved probe the energy landscape of our system and determine the activation energy (see Suppl. Info).

Finally, we functionalize the thermal dynamics by realizing with the skyrmion reshuffler device the key missing component for probabilistic computing using skyrmions. By assessing the device performance, we find an uncorrelated input and output signal with almost identical p-values. The errors of the obtained values are calculated as standard deviations. Their value is caused mainly by the statistical process of the skyrmion diffusion, as well as thermal movement of skyrmion in a non-flat potential landscape, where skyrmions have increased dwell time at distinct locations (see Suppl. Info). However, the decorrelation of the input and output signal is always strong even when calculating the highest possible deviation. The deviation of the p-value is caused by increased skyrmion dwelling at locations close to the threshold line for bit triggering. This is an effect of the different current densities in the reshuffling chamber and in the leading wire which can be overcome by optimizing the area for a better transition of skyrmions between these two sample regions. Our proof-of-concept devices demonstrates that the observed thermal skyrmion motion results in the creation of a signal highly uncorrelated to the device input while maintaining the same mean value of the telegraphic signal paving the way to implement skyrmion-based probabilistic computing.

## Methods

### Sample Parameters

The samples used in this study were grown as a continuous single layer stack of Ta(5)/Co$_{20}$Fe$_{60}$B$_{20}$(1)/Ta(0.08)/MgO(2)/Ta(5) (thickness in parentheses is given in nm). The sample was prepared by sputtering using a Singulus Rotaris sputtering tool with a base pressure $< 3 \times 10^{-8}$ mbar. Using a Singulus Rotaris deposition system, we can tune the thickness of the layers with high accuracy (reproducibility better than 0.01 nm). The stack was annealed at 200 °C for 1 hour to optimize the properties including the perpendicular magnetic anisotropy (PMA). It has been shown that introducing a very thin Ta interlayer between the CoFeB and MgO layers can change the PMA of the material and result in a skyrmion phase nucleation at room temperature[27]. By tuning the thicknesses and optimizing the sputter parameters we can obtain films that exhibit skyrmions at room temperature and by applying small out-of-plane (OOP) external fields, the skyrmion size can be tuned. We find at zero magnetic field that a stripe domain phase is present in the material. Within the resolution of the measurement, the ratio between the two magnetization states (up and down sweeps) is within the measurement precision 1:1, showing that the coercivity of the material is less than 0.05 mT. Samples were then patterned by electron beam lithography (EBL) into $60 \times 120\ \mu m^2$ pads with gold contacts for current injection. The sample was characterized with a superconducting quantum interference device (SQUID) and MOKE to determine the magnetic properties of the material stack. The magnetization loops are shown in Fig. 5.

The anisotropy field value was determined with hard axis loop measurement using SQUID, as in Ref. 28. As shown in Refs. 19,29, the DMI of a multi-stack material can be calculated from the domain pattern periodicity based on the measured values for anisotropy and saturation magnetization. We measured the worm domain periodicity around zero field and calculated the DMI using the obtained data assuming an exchange stiffness of $A = 10\ pJ\ m^{-1}$ [30]. The domain periodicity was evaluated using a fast Fourier transformation (FFT) of the MOKE pictures obtained at zero field. The material parameters at 300 K are summed up in Tab. 1.

Upon biasing with a current pulse, skyrmions already present in the sample are moved synchronously in a direction opposite to the current direction in line with the DMI and skyrmion Hall angle expected for a Ta-based stack[31]. The skyrmion velocity increases with current density and for $6.4 \times 10^{10}$ A m$^{-2}$ we obtain a velocity value of v = 5 mm s$^{-1}$, similar to what has been reported previously[18].

## Skyrmion Imaging and Tracking

We can use different approaches to nucleate skyrmions. The specific mechanism chosen for skyrmion generation is not crucial for our measurements, but we need to have a sufficiently low skyrmion density in order to prevent significant skyrmion-skyrmion interaction during the diffusion. While in our sample, skyrmions can for instance be nucleated by external field sweeps, however this does not allow for the skyrmion density to be easily controlled. A good control of the skyrmion density is achieved by nucleating skyrmions at the electric contacts by injecting electric current pulses. When applying a current density of $3.6 \times 10^{10}\ A\ m^{-2}$ for 5 ms, we have a 96 % probability of nucleating one skyrmion from the contact pad, see Suppl. Video 3. Different origins of the nucleation have been discussed, such as heating and spin torque effects in conjunction with local defects, as proposed in[32–35]. Our results indicate that a gradient in the current at the injection spot leads to skyrmion nucleation probably from a combination of heating and spin torques. As skyrmions are always nucleated at a certain location, we conclude the presence of a hotspot, acting as a nucleation site for the skyrmions during current injection. Such a hotspot is preferential since it allows for reproducible skyrmion injection along defined paths. This approach then allows us to fill the structure with skyrmions and due to their repulsive interaction, we find a rather homogeneous skyrmion density. An example of the sample filled with a relatively large number of skyrmions is shown in Fig. 6. The diffusion measurements are however carried out with lower skyrmion density to avoid skyrmion-skyrmion interaction.

We track the skyrmion motion with a 62.5 ms time resolution using the IMAGEJ[36]

software with the TRACKMATE plugin[37,38], see Suppl. Video 4. The identification of skyrmions is based on their contrast and intensity and the validity of the tracking algorithm was confirmed by manual analysis and comparison for a number of samples. In order to apply the diffusion theory and obtain a reliable diffusion coefficient, we have developed samples where skyrmions move the majority of the time. Therefore, we have concentrated here on samples, where we achieve skyrmion motion for more than 40 % of the measurement time to obtain robust diffusion coefficients with good statistics.

To make use of the full statistics, we use every skyrmion position as a reference point for that particular skyrmion tracking and a moving frame of reference analysis is employed[39]. This means that at lower times there is a significantly larger amount of data for analysis available and with increasing time the amount of data becomes smaller and the error bars increase. To compensate for this, we only consider the first half of the measurement time for the evaluation.

To change the sample temperature, a Peltier element QC-32-0.6-1.2 was used. It allows one to control the temperature of the sample with the sample temperature being measured using a Pt100 resistor. The stability of the set temperature is found to be within 0.3 K.

**Skyrmion Diffusion Theory**

We analyse the diffusion theory based on our atomistic spin model calculations. To reduce the computational effort, we use parameters corresponding to smaller skyrmions as the key diffusion properties are independent of the skyrmion size (see discussion in the main text). As with the Ta/Co20Fe60B20/Ta/MgO/Ta stack, our spin model is taylored to have a spin-spiral ground state below 35 K and supports the formation of isolated skyrmions with different topological charges at finite magnetic fields above 2 T. These skyrmions have a diameter on the order of few nanometre and are thus feasible to simulate on an atomistic level for sufficiently long timescale while this is challenging with the 1 µm-skyrmions in the CoFeB stacks used experimentally.

For the simulation of the skyrmion diffusion we use an atomistic spin Hamiltonian of the form

$$\mathcal{H} = -\frac{1}{2}\sum_{i,j} \mathbf{S}_i^\dagger \mathfrak{J}_{i,j} \mathbf{S}_j - \mathfrak{K} \sum_i S_{i,z}^2 - \mu_s \mu_0 \mathbf{H} \cdot \sum_i \mathbf{S}_i \\ -\frac{\mu_0 \mu_s^2}{4\pi} \sum_{i \neq j} \frac{3(\mathbf{S}_i \cdot \mathbf{r}_{ij})(\mathbf{S}_j \cdot \mathbf{r}_{ij}) - \mathbf{S}_i \cdot \mathbf{S}_j r_{ij}^2}{r_{ij}^5} \quad (4)$$

where $\mathbf{S}_i$ denotes the normalized magnetic moments of the Fe atoms. Here, we used the notation, in which the two-spin interaction $\mathfrak{J}_{ij}$ is a $3 \times 3$-matrix and includes the isotropic Heisenberg exchange $J_{ij} = \mathfrak{J}_{ij,\nu}^\nu/3$, as well as the DMI $D_{ij,\nu} = \varepsilon_{\alpha\beta\nu} \mathfrak{J}_{ij}^{\alpha\beta}/2$. These model parameters are similar to those of Tab.1 in Ref. 33, but rescaled with larger DMI. Furthermore, $\mathfrak{K}$ is the on-site anisotropy of 58.8 µRyd, $\mu_0 \mathbf{H}$ is the external magnetic field and $\mu_s = 3.3\,\mu_B$ is the magnetic moment of the Fe atoms. The last term in Eq. (4) are the long range dipolar interactions.

To investigate the skyrmion dynamics, we solve the stochastic LLG equation of motion[40]

$$\frac{\partial \mathbf{S}_i}{\partial t} = \frac{-\gamma}{(1+\alpha^2)\mu_s} \mathbf{S}_i \times (\mathbf{H}_i + \alpha \mathbf{S}_i \times \mathbf{H}_i) \quad (5)$$

where $\gamma = 1.76 \times 10^{11}\ T^{-1}s^{-1}$ is the gyromagnetic ratio and $\alpha$ the Gilbert damping constant. The effective field reads $\mathbf{H}_i = -\partial \mathcal{H}/\partial \mathbf{S}_i + \zeta_i$, where $\zeta_i$ are thermal fluctuations in the form of Gaussian white noise.

We start our simulations with a single skyrmion in the film centre and thermalize it at finite temperature for 5 ps ($\alpha = 1$) in the field polarized state ($\mu_0 H = 2.6$ T). The skyrmion is then traced over a time period of 5 ns, using an adaptation of the algorithm in Appendix A of Ref. 26, where the centre of magnetization $S_z(\mathbf{r})$ in a small region around the estimated skyrmion position $\mathrm{argmax}(S_z)$ is calculated (the ferromagnetic background is $S_z = -1$).

We use different methods to obtain the parameters of Eq. 2. The solid blue line in Fig. 2 represents a fit according to Eq. 2 yielding $\mathcal{D} = 5.12(59) \times 10^{-14}$ J s m$^{-2}$ and $\mathcal{G} = -3.43(17) \times 10^{-14}$ J s m$^{-2}$. The dashed red line represents the expected diffusive behavior following Eq. 2 for an ideal skyrmion, i.e. without thermal deformations of the spin structure. For this idealized situation, we obtain $\mathcal{D} = 5.0 \times 10^{-14}$ J s m$^{-2}$ and $\mathcal{G} = -3.35 \times 10^{-14}$ J s m$^{-2}$ for a phenomenological skyrmion profile

$$\boldsymbol{m}(\varrho) = -\mathrm{sech}[(\varrho - R)/\Delta]\mathbf{e}_\varrho - \tanh[(\varrho - R)/\Delta]\mathbf{e}_z, \tag{6}$$

with $R$ as radius and $\Delta$ as the skyrmion width fitted to the atomistic spin model. Both lines agree excellently and the diffusive motion of skyrmions is well described by Eq. 2. Calculating $\mathcal{D}$ and $\mathcal{G}$ directly from our simulations, including thermal distortions, yields a linearly decreasing dissipation and gyrocoupling for increasing temperatures.

**Supplementary Information**

**Skyrmion Diffusion dependence on temperature**

To understand the thermal dynamics better, we probe the key predictions of the analytical model for the diffusion coefficients: In the model the skyrmion diffusion coefficient is independent of the skyrmion size, but only dependent on the gyrocoupling and the friction (see Methods for the theoretical description). We measure the skyrmion size and resulting diffusion coefficient by tuning the size with the applied OOP external magnetic field. The threshold for the creation of skyrmion is at ~0.25 mT, forming skyrmions with radius of ca. 1 µm. The skyrmion size decreases linearly with larger field up to the ~0.40 mT. The diffusion coefficients depending on the skyrmion size are shown in Fig. 7. Smaller skyrmions exhibit a stronger diffusion, in contrast to the simple analytical model. To confirm this surprising finding, the measurement was repeated at a different temperature $T = 302$ K (see example as an inset in Fig. 7) and the same trend is found.

To check if the size dependence is a limitation of the analytical model, we next compare to numerical simulations. In the simulations we confirm the analytical prediction, that since dissipation and gyrocoupling are independent of the skyrmion's size, the purely diffusive motion is not expected to depend on the skyrmion's radius (see inset in Fig. 3).

The skyrmion radius $R$ at a given magnetic field is here calculated by fitting the $m_z$-component of Eq. (6). In contrast to experimental data, but in agreement with analytical expectations, the pure diffusive motion in a flat energy landscape does not depend on the size of the skyrmion in the simulations. So, from our measurements we can conclude that while skyrmions exhibit diffusive dynamics for the size range probed here, the experimentally observed diffusion exhibits an unexpected size dependence.

The second prediction that we probe is the predicted linear dependence of the diffusion coefficient on temperature. To check this, skyrmion diffusion was measured as a function of temperature using a MOKE microscope holder equipped with a Peltier element to control the temperature in the range of $270 - 330\,\text{K}$ (details see in the Methods section). The evaluation of the MSD dependence on time was done the same way as the measurement on room temperature. Examples of the MSDs and fitted diffusion slopes for three temperatures are shown in the inset in Fig. 8, highlighting that the pure diffusion behaviour holds for all probed temperatures.

The evaluated temperature dependence of the skyrmion diffusion coefficient is shown in Fig. 8. The size of the skyrmions was constant for a large part of the temperature range, only for the highest temperatures the skyrmion size slightly decreases. The lowest temperature that we can analyse for this sample stack is about $290\,\text{K}$ due to the time constraint of the total measurement time and the necessary displacement distances to obtain sufficient statistics for the low diffusion. The error of the diffusion coefficient was determined as a standard deviation of the linear fit slope calculated from the scatter of all MSD in one measurement. As the key observation, we do not find the expected linear dependence on temperature. We find rather a clear exponential diffusion coefficient dependence on temperature. Note that the temperature range is limited by the measurements in ambient atmosphere and above $305\,\text{K}$ more and more skyrmions are nucleated leading to a skyrmion lattice formation, indicating that the assumption of non-interacting skyrmions starts to break down.

Overall, we find good agreement between analytical theory, numerical simulations and the experiment for the MSD which is showing the same diffusive behaviour, the linear proportionality with time for all temperatures. However, for the skyrmion size and temperature dependence of the diffusion constant, clear differences are found, highlighting that the model by Schütte *et al.* put forward is not sufficient to capture the full physics. For systems completely homogeneous properties, the analytical model and the numerics agree in that there is no size dependence and further simulations show a linear temperature dependence of the diffusion coefficient in the range between 0 and 10 K (not shown) in line with the analytical theory. Furthermore, the results of the theoretical simulations are in very good agreement with the formula of Schütte, Eq. 2, and predict a room temperature diffusion coefficient in the range of $1.7 \times 10^{-9} m^2 s^{-1}$ and $9 \times 10^{-9} m^2 s^{-1}$ for typical damping parameters between $\alpha = 0.01$ and $\alpha = 0.05$. We now focus on the most striking difference between the experiments: We find that the experimental observations demonstrate that the skyrmion diffusion coefficient depends exponentially on temperature while theory (both analytics and numerics for homogeneous systems) predicts a linear dependence. To understand this difference, we plot in the inset of Fig. 9 the real-time data of the MSD for single skyrmions. The MSD, as calculated from the first frame of the video, evinces discrete jumps that skyrmions move from one point to another and stay there for some time. Such behaviour is known from diffusion in solids[41], where potential landscapes exist (in the simplest case given by the lattice). Our experimental observations show that we have also a non-flat potential landscape and we find that skyrmions have increased dwell times at certain randomly distributed positions. This observed behaviour actually allows us to use the dwell times to map an effective skyrmion pinning site distribution in the sample, as a major advantage of the result of the analysis of the thermally excited dynamics. Note that the model of diffusion stays applicable if the measurement time is much longer than such a dwell time at a single position, explaining why we find for long timescales good agreement with a pure diffusion dynamics model.

While simulations of a homogeneous system yield a linear temperature dependence of the skyrmion diffusion as expected from the analytical model (Eq. 2), simulations including a non-flat energy landscape, lead to an exponential dependence of the diffusion coefficient on temperature and thus fully reproduce the trend observed in the experiment. For the simulation of skyrmion diffusion with a non-flat energy landscape, a similar numerical setup was used: a skyrmion was placed in a lattice of locally varying magnetic properties with distances of 2.751 nm in x-direction and 4.7648 nm in y-direction (see Fig. 10), thermalized and tracked over a time period of 10 ns. The local variation that we implement at every of these points is to apply an additional field of -1.3 T out-of-plane to a cluster of five spins that represent one such localized position where skyrmions exhibit a larger dwell time. This effectively models a localized change of the magnetic moments of the material. The temperature dependence of the diffusion coefficient of a system with non-flat energy landscape is shown in Fig. 11. Within the temperature range of 8 K, the diffusion coefficient value changes over two orders of magnitude, yielding the same behaviour as observed experimentally.

Furthermore, this experimentally observed local dwelling of skyrmions at certain pinning sites also explains the lower absolute value of the diffusion coefficient as compared to the theoretical expectation for a homogeneous system. This is analogous to the change of the absolute diffusion values that go up significantly from the case of diffusion in solids, where pinning sites exist, to diffusion in liquids, where a flat energy landscape is effectively present[42]. This also naturally explains the measured exponential behaviour of the diffusion coefficient, thus resulting from the thermally activated depinning processes with an activation energy, which has been unknown so far. From the temperature dependence using an Arrhenius plot shown in Fig. 9, we can extract the average energy barrier for the skyrmions that we analyse experimentally and find a value of $E_A = 2.3(2) \times 10^{-19}\,J$. Overall, we see that skyrmions exhibit quasi-particle behaviour and they move as a whole, meaning that they largely act as rigid particles. Excitations of, for instance, only domain wall spin structure do not seem to play a significant role for the thermal dynamics.

The fact that thermal activation in inhomogeneous energy landscape leads to the observed behaviour also naturally explains the surprising size dependence of the skyrmion diffusion coefficient, which is not predicted by the simple theory. With larger skyrmion size, the area of magnetization opposite to the magnetization surrounding the skyrmion is larger. A larger skyrmion is therefore effectively more easily attracted to a pinning site (point with a localized potential minimum) than a skyrmion with smaller radius. The energy needed for a skyrmion stabilized at lower fields to be excited and exhibit diffusive motion is larger than for a skyrmion nucleated at higher field values.

By analysing the thermal motion, we explain the observed skyrmion behaviour as resulting from the superposition of pure diffusion and thermally activated hopping in an inhomogeneous energy landscape, showing that the absence of diffusive skyrmion dynamics in previous experiments can be attributed to strong pinning effects[43] and highlighting the necessity of full atomistic finite temperature simulations that goes beyond analytical theory of homogeneous systems to explain the experimentally observed dynamics.

# Acknowledgements

A. Donges and U. Nowak would like to acknowledge financial support by the DFG through SFB 767 at the University of Konstanz. M.K. and the group at Mainz acknowledge support by the DFG (in particular SFB TRR173 Spin+X), the Graduate School of Excellence Materials Science in Mainz (MAINZ, GSC 266) and WALL (FP7-PEOPLE-2013-ITN 608031). K. Everschor-Sitte and D. Pinna acknowledge financial support from the German Research Foundation (DFG) grant No. EV 196/2-1. J.Z. acknowledges the help and advice of the technicians of the Kläui group, especially S. Kauschke.

# Author contributions

M.K. and U.N. proposed and supervised the study. J.Z., S.J. and K.L. fabricated devices and characterized the multilayer samples. J.Z., D.H. prepared the measurement setup and together with N.K. and S.K. conducted the experiments using Kerr microscope. J.Z. and D.H. evaluated the experimental data with the help of P.V. and G.J., F.J. and A.D. performed the theoretical calculations and atomistic simulations of skyrmion diffusion. J.Z. produced, measured and analysed the skyrmion reshuffler under the supervision of D.P., K.E-S. and M.K. J.Z. drafted the manuscript with the help of M.K. and U.N. All authors commented on the manuscript.

## Data and code availability

The experimental data will be provided upon request by the corresponding author (klaeui@uni-mainz.de). The data from the simulation of skyrmion diffusion (and the custom code) can be provided upon request by Prof. Ulrich Nowak (ulrich.nowak@uni-konstanz.de).

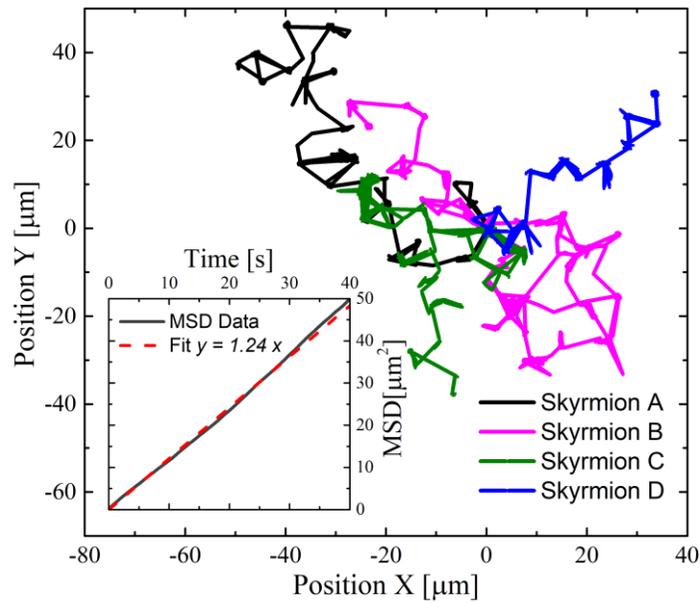

Fig. 1: Trajectories of selected skyrmions at 296 K. The reference position was taken from the first frame in the measurement. All skyrmions are set to start at position [0,0]. The timescale of the observation is in the seconds to minutes range. The inset shows MSD evaluated using the floating frame of reference, averaged over 15 skyrmions.

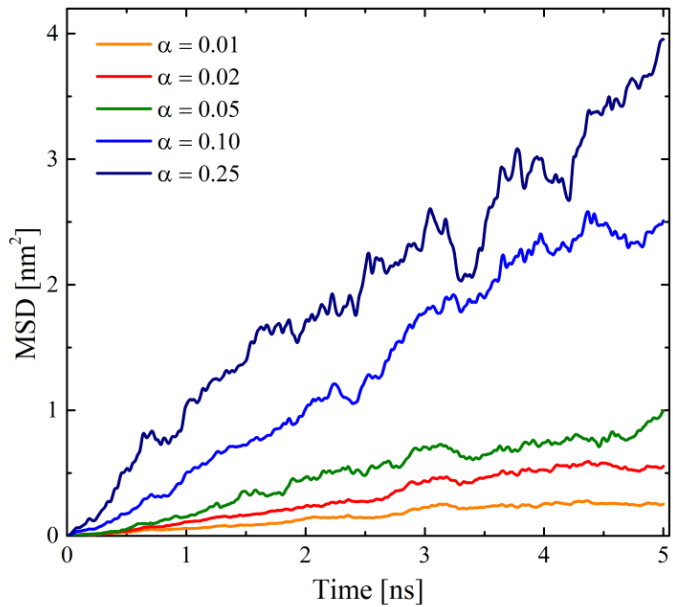

Fig. 2: Simulated dynamics of the MSD of single skyrmions for different damping values $\alpha$ showing diffusive behavior. For each curve 50 skyrmion trajectories were considered.

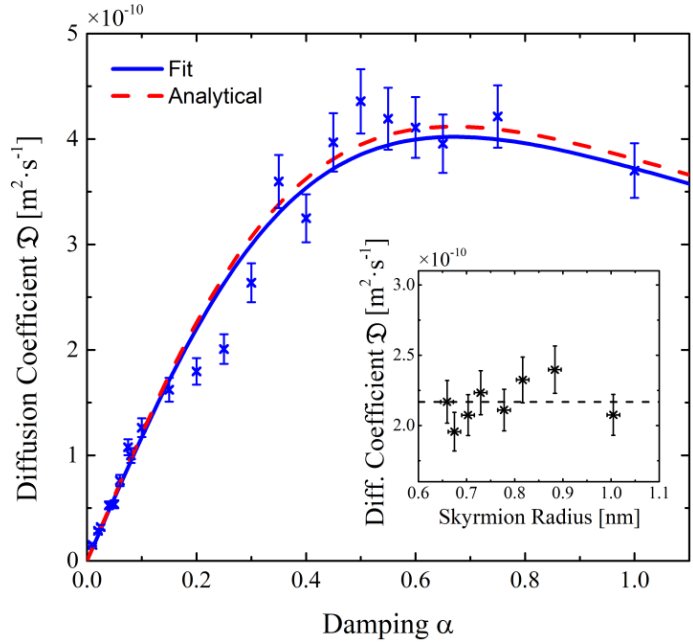

Fig. 3: Diffusion coefficient versus damping $\alpha$ at $T = 2\,K$ for simulated skyrmions. The solid line represents a fit after Eq. 2, while the dashed line corresponds to Eq. 2 with $\mathcal{D}$ and $\mathcal{G}$ calculated for an ideal skyrmion. The error is based on empirical variance and is calculated to be about $\pm 7\,\%$. Inset shows the diffusion coefficient versus skyrmion size for simulated skyrmions with damping parameter $\alpha = 0.2$. The dashed line represents the average of the calculated diffusion coefficients.

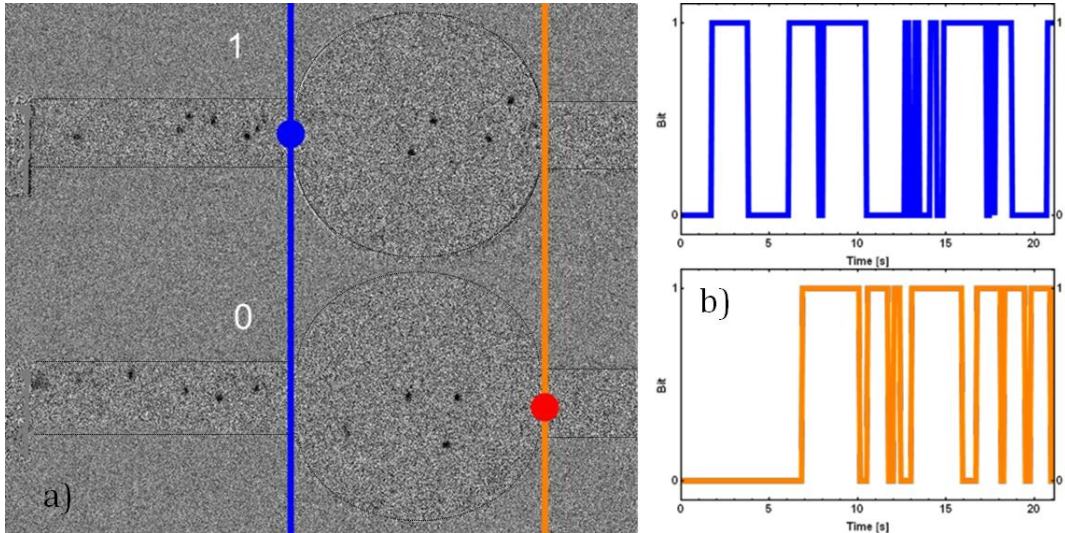

Fig. 4: a) Reshuffler operation with skyrmion nucleation by DC current. The input signal is constructed as a time frame where the skyrmion crosses the blue threshold line. The output is produced upon crossing the orange line. b) The corresponding input signal is depicted in blue colour on top and resulting output signal on the bottom. The full process of the reshuffler operation is shown in Suppl. Video 2. Radius of the reshuffling chamber is 80 μm. Reconstructed p-value for input signal is $p_{input} = 0.51 \pm 0.08$.

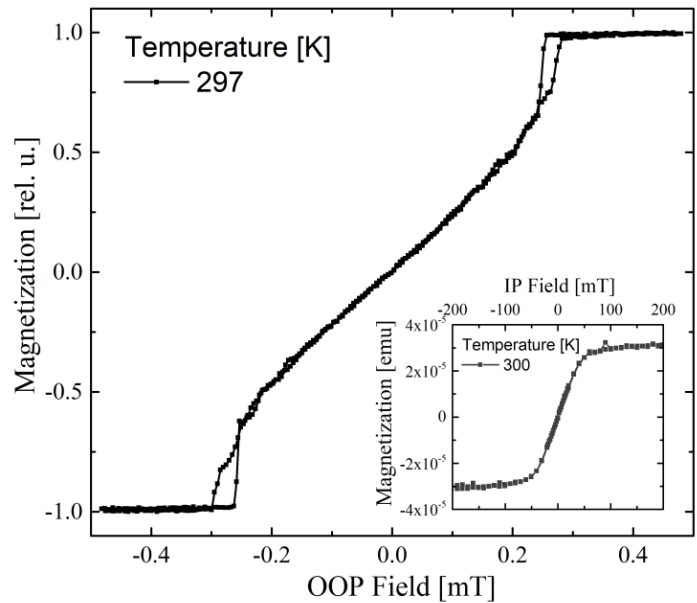

Fig. 5: Magnetization $M(B)$ loops with the external field perpendicular to the sample plane (out-of-plane configuration). The inset shows the magnetization loops with the field applied in the in-plane configuration.

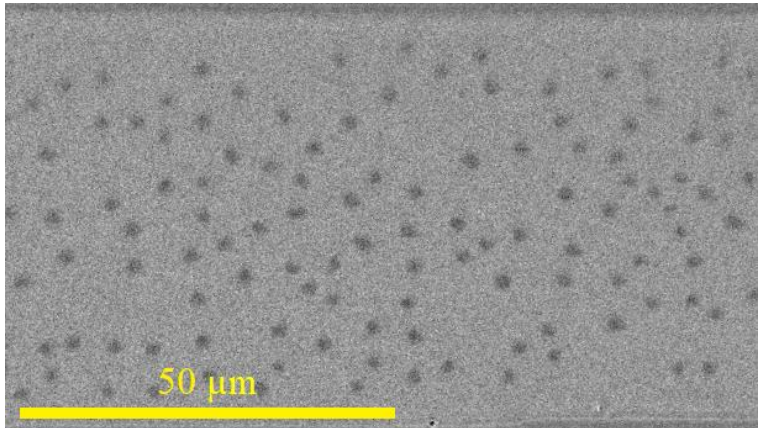

Fig. 6: Image of the sample with a large number of skyrmions injected. In this case, the distance between two skyrmions is in the range $5 - 8\,\mu m$ and the skyrmion density is relatively homogeneous. The diffusion in this state with a relatively high density of skyrmions was not evaluated as the movement in this state might be influenced with skyrmion-skyrmion interaction.

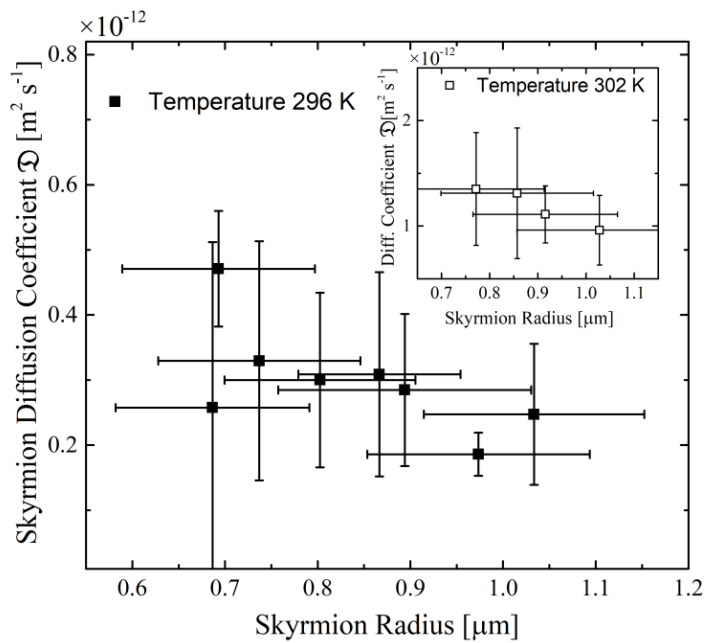

Fig. 7: Skyrmion diffusion coefficient as a function of the skyrmion radius at room temperature (296 K). The inset shows the radius dependence at a temperature of 302 K.

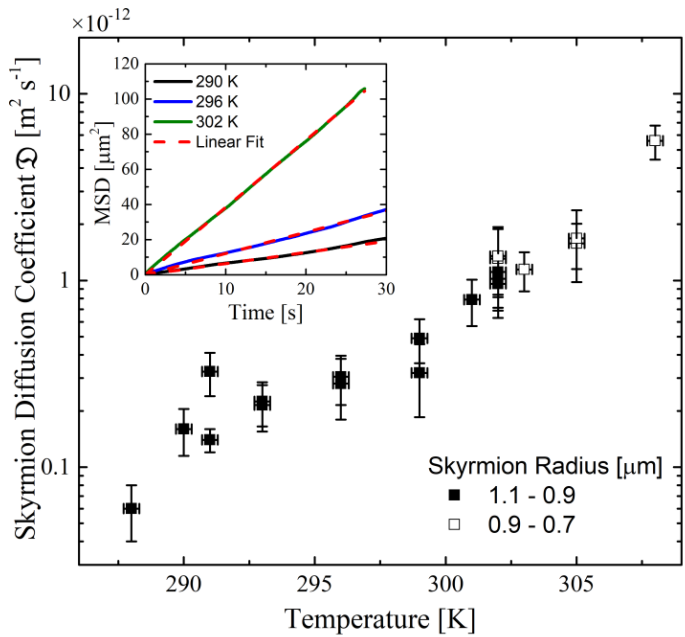

Fig. 8: Temperature dependence of the evaluated skyrmion diffusion coefficient. The linear dependence in the logarithmic plot shows a clear exponential dependence of the diffusion coefficient on temperature over 2 orders of magnitude. The inset shows examples of the MSD and their linear fits for different temperatures.

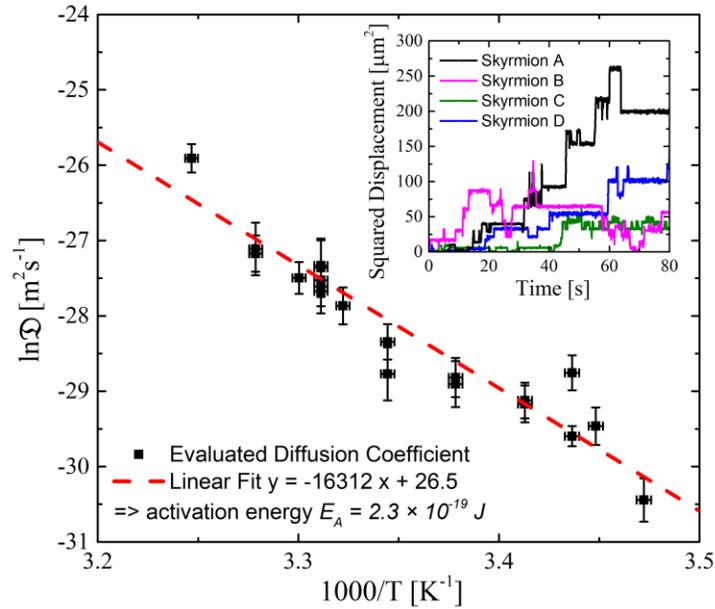

Fig. 9: The Arrhenius plot of the calculated skyrmion diffusion coefficients. Inset depicts the trajectories of selected skyrmions. The reference position was taken from the first frame in the measurement.

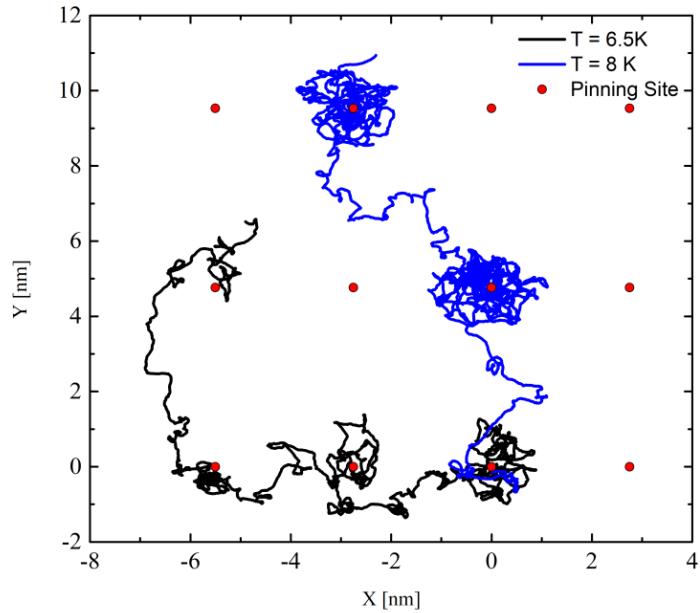

Fig. 10: Simulation of skyrmion diffusion with locally varying magnetic properties that act as pinning sites in a lattice. The skyrmion exhibits an increased dwell time at the pinning sites and the trajectory resembles the experimentally observed trajectories.

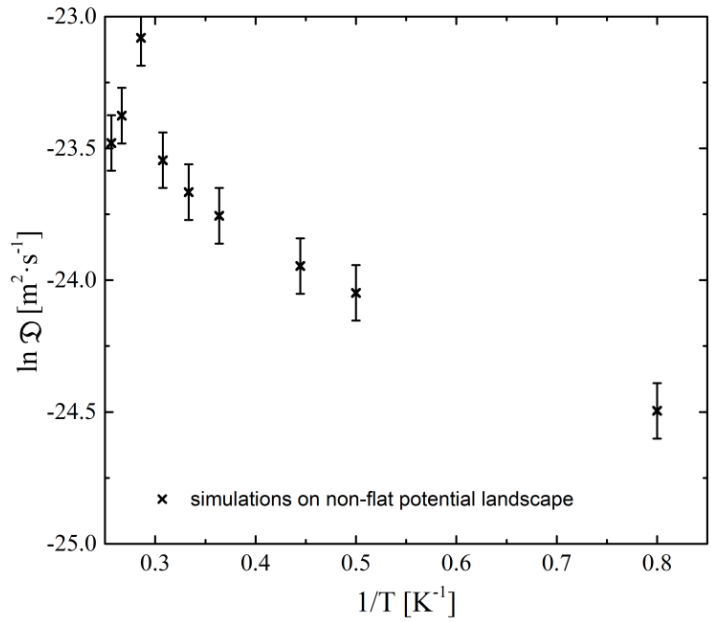

Fig. 11: Temperature dependence of the skyrmion diffusion coefficient extracted from simulations with a non-flat energy potential landscape, using the damping parameter $\alpha = 0.4$. The error is based on empirical variance and is found to be about $\pm 7\ \%$.

Table 1: Evaluated saturation magnetization $M_S$, anisotropy field $\mu_0 H_K$, Dzyaloshinskii-Moriya interaction $D$ and stripe periodicity $\lambda$ of the used materials stack.

| $M_S$(A m$^{-1}$) | $\mu_0 H_K$(mT) | $D$(mJ m$^{-2}$) | $\lambda$(μm) |
|---|---|---|---|
| $7.36(5) \times 10^5$ | 67(6) | 0.30(8) | 3.2(3) |

**Supplementary Video Captions**

Suppl. Video 1: Skyrmion motion in a relaxed state at temperature 296 K in a constant 0.35 mT out-of-plane field. No current or field pulses are applied.

Suppl. Video 2: Operation of the skyrmion reshuffler device upon application of DC current. Skyrmions are continuously generated at the gold contact pad and transported through the device while undergoing diffusive motion. The input and output signal are constructed by skyrmion crossing the over threshold line and triggering the corresponding bit into the input for the blue line and into the output for the orange line, respectively. Current density is set as $j = 3 \times 10^8 \, Am^{-2}$ and out-of-plane magnetic field at 0.2 mT. Field of view of the video is 165×135 μm².

Suppl. Video 3: Skyrmion nucleation with current pulses. The used current density of $3.6 \times 10^{10}$ A m$^{-2}$ is applied for 5 ms with a frequency of 2 Hz. One current pulse results in a 96 % probability of nucleating one skyrmion from the contact pad.

Suppl. Video 4: Tracking of five selected skyrmions using the ImageJ software with the TrackMate plugin[37,38]. The measurement is done at temperature 296 K in a constant 0.35 mT out-of-plane field with a 62.5 ms time resolution. Video dimensions are $178 \times 135$ μm².